\documentclass[12pt]{article}
\usepackage{graphicx}
\def\Journal#1#2#3#4{{#1} {\bf #2}, #3 (#4)}

\def\PLB{Phys. Lett. B}

\def\PRD{Phys. Rev. D}
\def\ZPC{Z. Phys. C}
\def\EPJ{Eur. Phys. J. C}

\newcommand{\ra}{\rightarrow}
\newcommand{\Lam}{\Lambda_{c}^{+}}
\newcommand{\Lambar}{\bar{\Lambda}_{c}^{+}}
\newcommand{\Lamvec}{\vec{\Lambda}_{c}^{+}}
\newcommand{\Lambarvec}{\vec{\bar{\Lambda}}_{c}^{{}_{{}_+}}}

\begin{document}

\begin{center}
\baselineskip=1cm
{\Large \bf Charmed Hadron Production \\
in Neutrino Reactions \\
and Polarized $s$--quark Distribution\footnote[1]{Talk presented by 
K. Sudoh at the NuFACT'01 workshop, Tsukuba, Japan, May 24--30 (2001).}} 

\baselineskip=0.6cm
\vspace{1.5cm}
Kazutaka ~SUDOH \\
\vspace{0.2cm}
{\em Graduate School of Science and Technology, \\ 
Kobe University, Nada, Kobe 657--8501, Japan} \\
{\tt E-mail:sudou@radix.h.kobe-u.ac.jp}

\vspace{0.8cm}
Toshiyuki ~MORII \\
\vspace{0.2cm}
{\em Faculty of Human Development, \\
Kobe University, Nada, Kobe 657--8501, Japan} \\
{\tt E-mail:morii@kobe-u.ac.jp}
\end{center}

\vspace{1.0cm} 
\noindent
\begin{center}
{\bf Abstract}
\end{center}

In order to obtain information about the polarized strange quark 
distribution, we studied the semi--inclusive $\Lam$/$\Lambar$ production 
in neutrino and polarized proton collisions. 
We found that these reactions are effective to clearly extract the 
polarization of strange quark by measuring the spin correlation between 
the target proton and the produced $\Lam$/$\Lambar$ baryon. 

\clearpage

Proton spin puzzle is currently one of the most challenging topics 
in high energy spin physics. 
As is well known, proton spin is composed of the spin and orbital 
angular momentum of quarks and gluons which constitute proton. 
The polarized parton distribution plays an important role on 
the spin structure of proton. 
However, knowledge of the polarized sea quark distribution remain still 
poor. 
In order to understand the spin structure of proton, we need more 
information about the polarized sea quark distribution functions. 
Recently, HERMES group has reported \cite{HERMES99} that direct 
measurement of the strange sea is required to explain the violation of 
the Ellis--Jaffe sum rule \cite{Ellis74}. 
So far, there are several parametrization models of the strange quark 
distribution. 
Though the most simplest case is to assume the flavor ${\rm SU(3)}_f$, 
a new parametrization taken account of the violation of ${\rm SU(3)}_f$ 
is also recently proposed \cite{Leader99}. 

Here we focus on the polarized $s$/$\bar{s}$ quark distribution. 
To examine the polarized $s$/$\bar{s}$ quark distribution, we have 
studied semi--inclusive $\Lam$/$\Lambar$ neutrino production; 
$\nu + \vec{p} \ra l^- + \Lamvec + X$, 
$\bar{\nu} + \vec{p} \ra l^+ + \Lambarvec + X$, 
which might be observed at the planned factory, where arrows attached 
to particles mean that these particles are polarized. 
Since in the naive quark model, the $\Lam$ baryon is composed of a heavy 
$c$--quark and anti--symmetrically combined light $u$ and $d$--quarks, 
we can assume the polarization of $\Lam$ baryon to be the one of $c$--quark. 
In addition, $\Lam$ is dominantly produced by the fragmentation of 
$c$--quark which is originated from $s$--quark through the $t$--channel 
$W$ exchange at the leading order. 
Therefore, there can be a correlation between the $s$--quark 
polarization and the produced $c$--quark polarization. 
Hence we can expect that the measurement of the spin correlation 
between the incident proton and the produced $\Lam$/$\Lambar$ gives us 
information about the polarized $s$/$\bar{s}$--quark distribution in proton. 

For above processes, we calculated the double spin asymmetry $A_{LL}^H$ 
for final state hadron specified by $H$ ($\Lam$ or $\Lambar$), 
which is given by 
\begin{equation}
A_{LL}^{H}=\frac{d\Delta\sigma/dp_{T}}{d\sigma/dp_{T}} ,
\label{ALL}
\end{equation}
where $p_T$ is a transverse momentum of final hadron $H$. 
$d\Delta\sigma$ represents the spin--dependent differential cross 
section and is defined in terms of $d\sigma_{h h^{\prime}}$ as 
\begin{equation}
d\Delta\sigma \equiv \frac{1}{4} [ d\sigma_{++} - d\sigma_{+-}
+ d\sigma_{--} - d\sigma_{-+}] ,
\end{equation}
with definite helicities $h$ and $h^{\prime}$ for incoming 
proton and outgoing $\Lam$/$\Lambar$, respectively. 

For $\Lam$ production, the spin--dependent differential 
cross section is given 
\begin{equation}
d\Delta\sigma(\nu p\ra l^{-}\Lam X)
=\left\{{\rm U}_{cs}^2 \Delta s(x)+{\rm U}_{cd}^2 \Delta d(x)\right\}dx
\left(\frac{d\Delta\hat{\sigma}}{d\hat{t}}
\right)d\hat{t}\Delta D_c^{\Lam} (z)dz ,
\end{equation}
where $\Delta s(x)$ and $\Delta d(x)$ are the polarized 
$s$--quark and $d$--quark distribution functions, respectively.
${\rm U}_{cs}$ and ${\rm U}_{cd}$ are CKM parameters. 
$\Delta D_c^{\Lam}(z)$ is the polarized fragmentation function of 
outgoing charm quark decaying into $\Lam$. 
We used the model of Peterson {\em et al.} \cite{Peterson83} as the 
unpolarized fragmentation function, which is given by \cite{PDG98} 
\begin{equation}
D_c^{\Lam}(z)=\frac{1}{z[1-\frac{1}{z}-\frac{\epsilon_p}{1-z}]^2}
~~~~~~~~~\epsilon_p\sim 0.25~~\mbox{for}~~\Lam . 
\end{equation}
Unfortunately, the polarized fragmentation function is not yet 
established because of lack of experimental data. 
By analogy with the study on $\Lambda$ polarization \cite{deFlorian98}, 
we took the 
following ansatz: 
\begin{equation}
\Delta D_c^{\Lam}(z)=C_c^{\Lam}(z) D_c^{\Lam}(z) , 
\end{equation}
where $C_c^{\Lam}(z)$ is the scale--independent spin transfer 
coefficient. 
Here we apply the analysis on $\Lambda$ production to $\Lam$ production 
and choose the following two models; $C_c^{\Lam}(z)=1$ (the naive 
nonrelativistic quark model) and $C_c^{\Lam}(z)=z$ (the jet 
fragmentation models \cite{Bartl80}). 
The double spin asymmetry $A_{LL}^{\Lam}$ is described in terms of 
the spin transfer coefficient $C_c^{\Lam}(z)$ and the ratio of the 
parton distribution functions as 
\begin{equation}
A_{LL}^{\Lam}\propto C_c^{\Lam}(z)\frac{\left\{{\rm U}_{cs}^2 \Delta s(x)
+{\rm U}_{cd}^2 \Delta d(x)\right\}}{\left\{{\rm U}_{cs}^2 s(x)
+{\rm U}_{cd}^2 d(x)\right\}}. 
\end{equation}
Thus $A_{LL}^{\Lam}$ is proportional to linear combination of 
the polarized $s$--quark and $d$--quark distribution function. 
Therefore, $\Lam$ production is not so good process for clearly 
extracting the $s$--quark distribution, since the contribution from 
the valence $d$--quark is large. 

On the contrary, $A_{LL}^{\Lambar}$ in $\Lambar$ production can be
represented as 
\begin{equation}
A_{LL}^{\Lambar}\propto C_c^{\Lam}(z)\frac{\left\{{\rm U}_{cs}^2 \Delta 
\bar{s}(x)+{\rm U}_{cd}^2 \Delta \bar{d}(x)\right\}}
{\left\{{\rm U}_{cs}^2 \bar{s}(x)+{\rm U}_{cd}^2 \bar{d}(x)\right\}} 
=C_c^{\Lam}(z)\frac{\Delta \bar{s}(x)}{\bar{s}(x)}
=C_c^{\Lam}(z)\frac{\Delta s(x)}{s(x)}, 
\end{equation}
in which the $\bar{d}$--quark contribution can be eliminated and thus 
$A_{LL}^{\Lambar}$ is directly proportional to the strange quark 
distribution function. 
Therefore, we can clearly extract the polarized strange quark 
distribution $\Delta s(x)$. 
Note that above equation is derived in both the flavor ${\rm SU(3)}_f$ 
case ($\Delta \bar{u}(x)=\Delta \bar{d}(x)=\Delta \bar{s}(x)$)
and the non--${\rm SU(3)}_f$ case 
($\Delta \bar{u}(x)=\Delta \bar{d}(x)=\lambda \Delta \bar{s}(x)$\footnote
{$\lambda$ represent a degree of ${\rm SU(3)}_f$ violation and is a 
parameter which should be determined from experiments \cite{Leader99}.}). 

Setting a charm quark mass $m_c = 1.5$ GeV and the relevant collider 
energy $\sqrt{s}=50$ GeV, we numerically calculated the 
spin--independent and dependent differential cross sections and the 
double spin asymmetry. 
As for the parton distribution functions, we used the GRV98 \cite{GRV98} 
parametrization as the unpolarized parton distribution function, 
and the AAC \cite{AAC} and ``standard scenario'' of 
GRSV00 \cite{GRSV00} parametrizations for the polarized one. 

\begin{figure}[t!]
 \begin{center}
   \includegraphics[width=16pc, keepaspectratio]{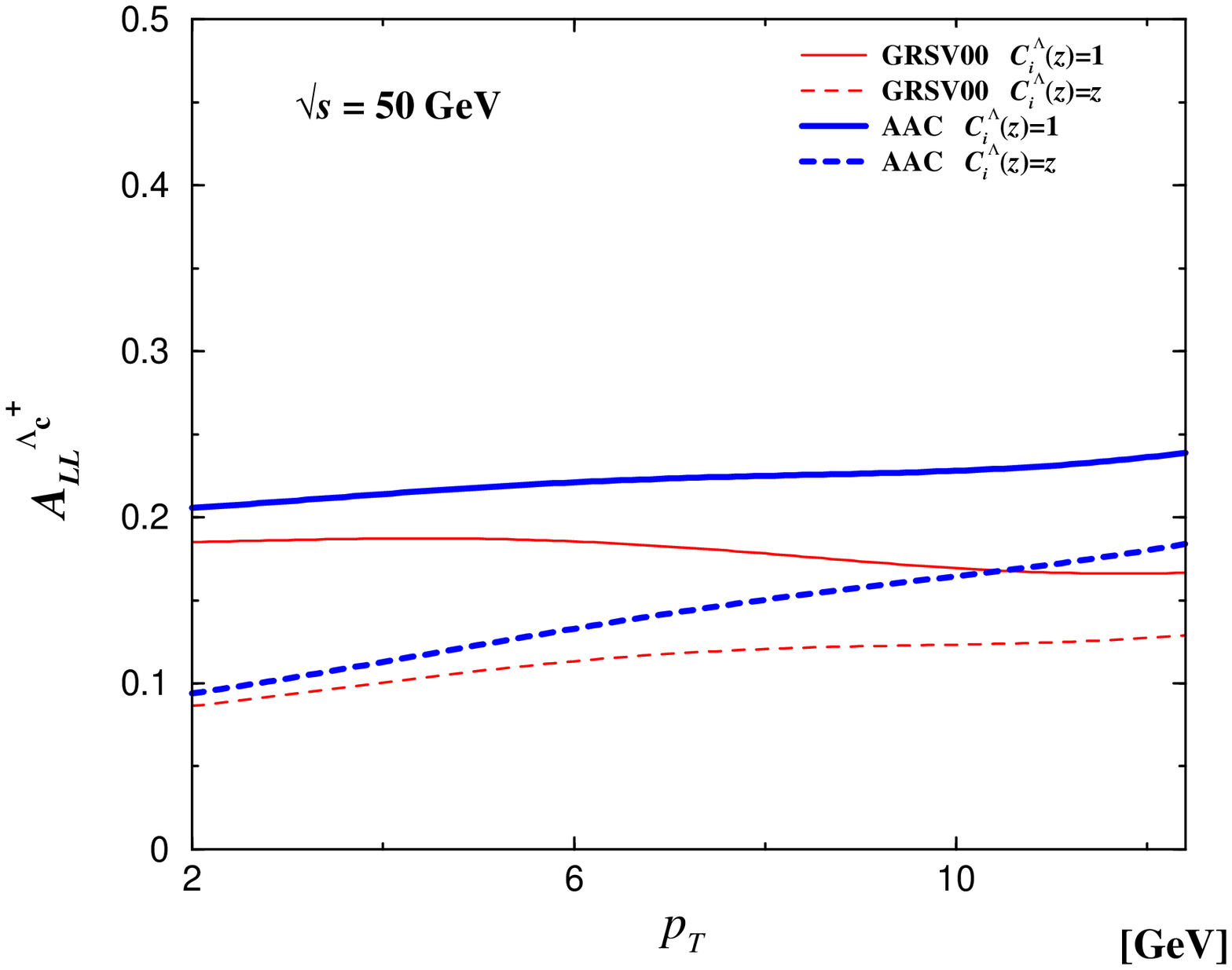}
   \includegraphics[width=16pc, keepaspectratio]{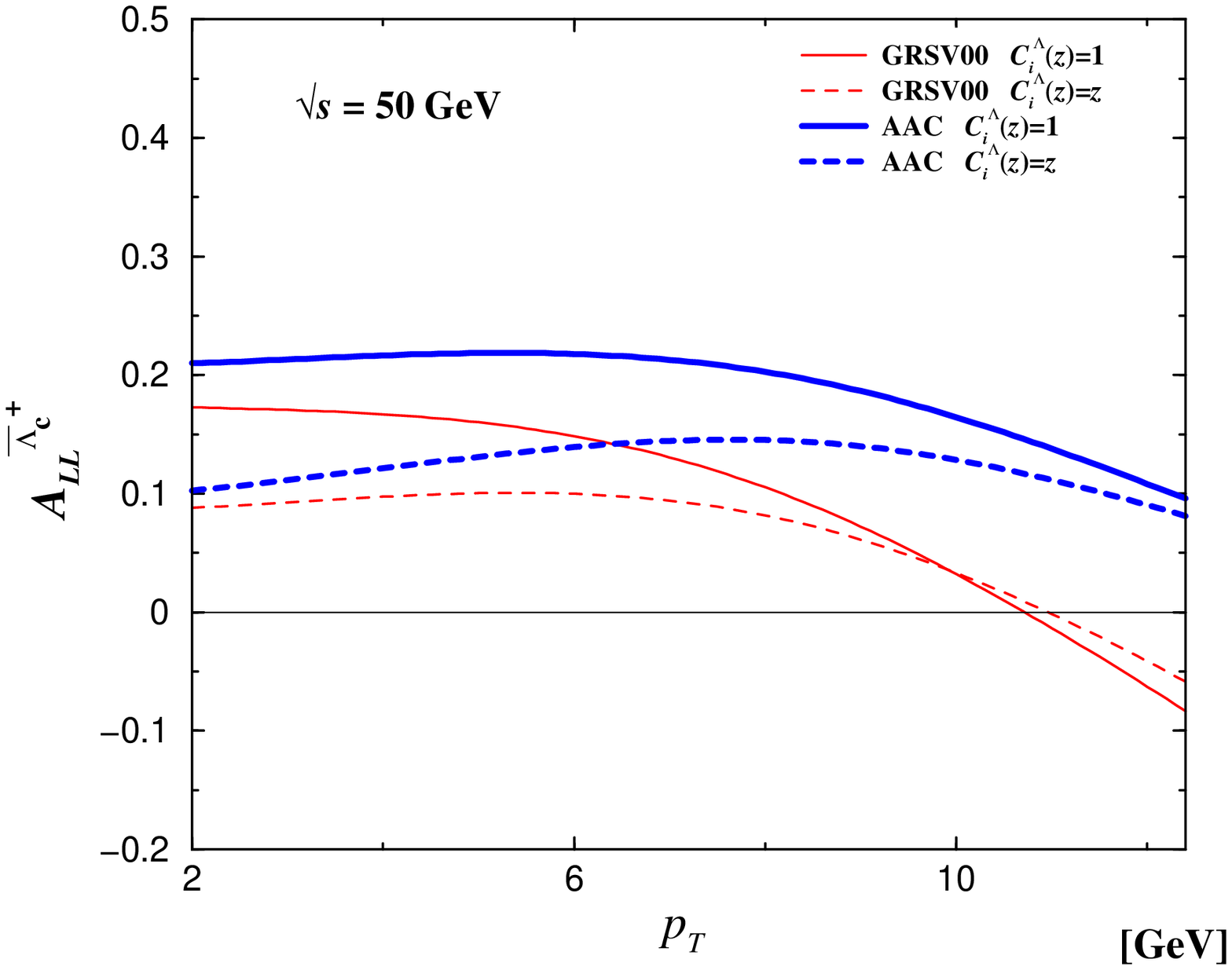}
 \end{center}
\caption{
$p_T$ distribution of double spin asymmetries at $\sqrt{s}=50$ GeV
for $\Lam$ production (left panel) and 
for $\Lambar$ production (right panel). 
}
\label{fig1}
\end{figure}

We show the double spin asymmetry in Fig. \ref{fig1} as a function of 
$p_T$ at $\sqrt{s}=50$ GeV for $\Lam$ production (left panel) and for 
$\Lambar$ production (right panel). 
In order to suppress the contributions from the diffractive process and 
higher twist corrections, we have imposed the kinematical cut on $p_T$ 
as $p_T > 2$ GeV in numerical calculations. 
In figures, the bold and normal lines show the case of AAC 
parametrization and ``standard scenario'' of GRSV00 parametrization, 
respectively. 
The solid lines represent the spin transfer coefficient 
$C_c^{\Lam}(z)=1$ case, while the dashed lines represent 
$C_c^{\Lam}(z)=z$ case. 

As shown in figures, $A_{LL}^{\Lam}$ in the smaller $p_T$ regions does 
not depend on the model of polarized parton distribution functions, 
and is strongly affected by the shape of the spin transfer coefficient 
$C_c^{\Lam}(z)$. 
Therefore, The $\Lam$ production is effective for extracting information 
about the polarized fragmentation function $\Delta D_c^{\Lam}(z)$. 
On the other hand, measuring $A_{LL}^{\Lambar}$ in larger $p_T$ regions 
is quite effective for testing the model of polarized parton 
distribution functions, since the ambiguity of the polarized 
fragmentation function is small and we see a big difference between two 
parametrization models. 

In summary, to extract information about the polarized strange quark 
distribution in proton, the semi--inclusive $\Lam$/$\Lambar$ production 
in neutrino--polarized proton collisions was studied. 
$\Lambar$ production is most promising not only for testing but also 
for directly extracting the polarized strange quark distribution 
$\Delta s(x)$ by measuring $A_{LL}^{\Lambar}$. 

\begin{thebibliography}{00}
\bibitem{HERMES99}
HERMES Collaboration, K. Ackerstaff {\em et al.}, 
\Journal{\PLB}{464}{123}{1999}. 
\bibitem{Ellis74}
J.Ellis and R. L. Jaffe, 
\Journal{\PRD}{9}{1444}{1974}; \Journal{\PRD}{10}{1669}{1974}. 
\bibitem{Leader99}
E. Leader, A. V. Sidorov, D. B. Stamenov, 
\Journal{\PLB}{462}{189}{1999}. 
\bibitem{Peterson83}
C. Peterson, D.Schlatter, I. Schmitt, P. M. Zerwas, 
\Journal{\PRD}{27}{105}{1983}. 
\bibitem{PDG98}
Particle Data Group, C. Coso {\em et al.}, 
\Journal{\EPJ}{3}{1}{1998}. 
\bibitem{deFlorian98}
D. de Florian, M. Stratmann, W. Vogelsang, 
\Journal{\PRD}{57}{5811}{1998}. 
\bibitem{Bartl80}
A. Bartl, H. Fraas, W. Majerotto, 
\Journal{\ZPC}{6}{335}{1980}. 
\bibitem{GRV98}
M. Gl\"uck, E. Reya, A. Vogt, 
\Journal{\EPJ}{5}{461}{1998}. 
\bibitem{AAC}
Asymmetry Analysis Collaboration, Y. Goto {\em et al.}, 
\Journal{\PRD}{62}{034017}{2000}. 
\bibitem{GRSV00}
M. Gl\"uck, E. Reya, M. Stratmann, W. Vogelsang, 
\Journal{\PRD}{63}{094005}{2001}. 
\end{thebibliography}
\end{document}